\documentclass[acmsmall]{acmart}
\AtBeginDocument{%
  }

\usepackage{booktabs}
\usepackage{setspace} 
\usepackage{amsmath}
\usepackage{xurl}
\usepackage[most]{tcolorbox}
\usepackage{xspace}
\usepackage{multicol}
\usepackage{multirow}
\usepackage{tabularx}
\usepackage{array}   

\usepackage[dvipsnames,table,xcdraw]{xcolor}
\usepackage{enumitem}
\usepackage{subcaption}
\usepackage{graphicx}
\usepackage{comment}
\usepackage{xparse}
\usepackage{mathtools}
\usepackage{ragged2e}
\usepackage{amsthm}
\usepackage{microtype}  
\usepackage{scalerel}
\usepackage{bm}

\usepackage{amsfonts}
\usepackage{pifont}
\usepackage[normalem]{ulem}
\usepackage[strings]{underscore}
\usepackage{epigraph}
\usepackage{color,soul}
\usepackage{longtable}

\usepackage{csquotes}
\usepackage{tikz}
\usepackage{venndiagram}
\usepackage{calc}
\usepackage{fp}
\newcounter{z}
\usepackage{svg}
\usepackage{float}
\usepackage{fontawesome5}

\usepackage{listings}

\usepackage{listings}

\renewcommand{\subparagraph}[1]{
\vspace{2mm}
\textit{#1}
}

\newcommand{\bnm}{\begin{newmath}}
\newcommand{\enm}{\end{newmath}}

\newcommand{\bea}{\begin{neweqnarrays}}%
\newcommand{\eea}{\end{neweqnarrays}}%

\newcommand{\bne}{\begin{newequation}}
\newcommand{\ene}{\end{newequation}}

\newcommand{\bal}{\begin{newalign}}
\newcommand{\eal}{\end{newalign}}

\newtcolorbox{cutegreenbox}[1]{colback=green!5,colframe=green!35!black,fonttitle=\bfseries,title={#1}}

\newcommand{\naman}[1]{\textbf{{\color{teal}\emph{NG:~ #1}}}}

\definecolor{DarkGreen}{RGB}{1,50,32}

\newcommand{\lgbt}{LGBTQ2S+\xspace}

\newcommand{\TFA}{Technology-Facilitated Abuse}

\newcommand{\univ}{[U.S. University]\xspace}  
\newcommand{\wi}{[US. State]\xspace}  
\newcommand{\techclinic}{[Tech Clinic]\xspace} 
\newcommand{\hochunk}{[Tribe]\xspace}
\newcommand{\lco}{[Tribe]\xspace}
\newcommand{\teejop}{[City]\xspace}
\newcommand{\was}{[National-level Indigenous Conference]\xspace}
\newcommand{\mend}{[State-level Organization (Native)]\xspace}
\newcommand{\arc}{[Local-level Organization]\xspace}

\newlist{researchquestions}{enumerate}{1}
\setlist[researchquestions]{label*=\textbf{RQ\arabic*}}

\newcounter{toplvl}
\newcounter{sublvl}[toplvl]  

\newlist{numlist}{itemize}{1}
\setlist[numlist]{label=, leftmargin=1.3em, itemindent=0pt, labelsep=0.5em, itemsep=0.2em}

\newcommand{\tlevel}[1]{%
  \setcounter{toplvl}{#1}%
  \setcounter{sublvl}{0}%
}

\newcommand{\slevel}{%
  \stepcounter{sublvl}%
  \item[\thetoplvl.\thesublvl]%
}

\usetikzlibrary{shapes,positioning,calc}

\newcommand{\tstat}[3]{%
    ($#1$, $T=#2$, $p=#3$)
}

\newcommand{\ntraining}{8}
\newcommand{\ntotal}{189}
\newcommand{\nsurvey}{68}
\newcommand{\npre}{57}
\newcommand{\nnoconsent}{6}
\newcommand{\npost}{36}
\newcommand{\ndemographics}{24}
\newcommand{\ncommon}{16}
\newcommand{\nfollowup}{7}
\newcommand{\nlesson}{3}
\newcommand{\nchecklist}{24}

\newcommand\varpm{\mathbin{\vcenter{\hbox{%
  \oalign{\hfil$\scriptstyle+$\hfil\cr
          \noalign{\kern-.3ex}
          $\scriptscriptstyle({-})$\cr}%
}}}}
\newcommand\varmp{\mathbin{\vcenter{\hbox{%
  \oalign{$\scriptstyle({+})$\cr
          \noalign{\kern-.3ex}
          \hfil$\scriptscriptstyle-$\hfil\cr}%
}}}}

\definecolor{headerDark}{HTML}{2C3E50}      
\definecolor{headerWarm}{HTML}{8B1A1A}      
\definecolor{rowAlt}{HTML}{F5F0EB}          
\definecolor{rowBase}{HTML}{FFFFFF}         
\definecolor{labelCol}{HTML}{1C1C1C}        
\definecolor{accentRule}{HTML}{8B1A1A}      
\tcbset{
  badgebase/.style={
    boxrule=0.3pt,
    arc=1pt,
    boxsep=0.5pt,
    left=1pt,
    right=1pt,
    top=1pt,
    bottom=1pt,
    fontupper=\bfseries,
    on line
  },
  goalbadge/.style={
    badgebase,
    colback=teal!10,
    colupper=teal!50!black,
    colframe=teal!70
  },
  purplebadge/.style={
    badgebase,
    colback=purple!10,
    colupper=purple!50!black,
    colframe=purple!70
  },
  recbadge/.style={
    badgebase,
    colback=violet!15,
    colupper=violet!50!black,
    colframe=violet!40
  }
}

\newcommand{\badge}[4]{%
  \tcbox[#1]{%
    \hyperref[#2]{\textcolor{#3}{#4}}%
  }%
}

\newcommand{\goal}[1]{%
  \badge{goalbadge}{tab:evaluation-goals}{teal!50!black}{EG#1}%
  \label{tab:goal-EG#1}%
}

\newcommand{\goalref}[1]{%
  \badge{goalbadge}{tab:goal-EG#1}{teal!50!black}{EG#1}%
}

\newcommand{\objective}[1]{%
  \badge{purplebadge}{tab:usefulness}{purple!50!black}{O#1}%
  \label{O-#1}%
}

\newcommand{\objref}[1]{%
  \badge{purplebadge}{O-#1}{purple!50!black}{\texttt{O#1}}%
}

\newcommand{\lesson}[1]{%
  \badge{purplebadge}{tab:lessons}{purple!50!black}{\texttt{L#1}}%
  \label{tab:lesson-#1}%
}

\newcommand{\lessonref}[1]{%
  \badge{purplebadge}{tab:lesson-#1}{purple!50!black}{\texttt{L#1}}%
}

\providecommand{\setstretch}[1]{}

\newcommand{\mycaption}[2]{
    \caption[#1]{\textbf{#1} -- {\small\textmd{#2}}}
}

\newenvironment{myquote}[1]{%
  \def\quoteauthor{#1}
  \begin{list}{}{%
    \leftmargin0.3cm
    \rightmargin\leftmargin
  }%
  \item\relax
  \itshape ``%
}{
  ’’
  \hspace{0.2cm}\normalfont\small\hfill{--- \quoteauthor}%
  \end{list}%
}

\newenvironment{myquotel}[1]{%
  \def\quoteauthor{#1}
  \begin{list}{}{%
    \leftmargin0.3cm
    \rightmargin\leftmargin
  }%
  \item\relax
  \itshape``
}{
  ''
  \par\vspace{-0.5ex}
  \normalfont\small\hfill--- \quoteauthor%
  \end{list}%
}

\newcommand{\inline}[1]{``\textit{#1}''}

\newcommand{\graycell}{\cellcolor{gray!15!white}}

\newcommand{\fixme}[1]{{\color{Red}{[\textbf{FIXME: #1}]}}}

\def\authnote{1}
\newcounter{rcnote}[section]
\newcommand{\rcnote}[1]{\ifnum\authnote=1\refstepcounter{rcnote}{\bf \textcolor{magenta}{$\ll$RC: {\sf #1}$\gg$}}\fi}

\newcommand{\tabfontsize}{\footnotesize}

\newcommand{\bulletpoint}{$\lozenge$\xspace }

 \newcounter{tablerow}

\NewDocumentCommand{\pickoneitem}{m}{\item #1}

\NewDocumentCommand{\pickone}{m+m+m+g}{%
  \textbf{\color{DarkGreen}\emph{PICK ONE:}}%
  \begin{itemize}
    \pickoneitem{#1}%
    \pickoneitem{#2}%
    \pickoneitem{#3}%
    \IfNoValueF{#4}{\pickonemore#4}%
  \end{itemize}
}

\NewDocumentCommand{\pickonemore}{m+g}{%
  \pickoneitem{#1}%
  \IfNoValueF{#2}{\pickonemore#2}%
}

\definecolor{acm-purple}{RGB}{90, 53, 158}
\definecolor{acm-green}{RGB}{0, 102, 0}
\definecolor{acm-darkblue}{RGB}{0, 51, 102}

\hypersetup{
  colorlinks   = true,
  linkcolor    = acm-purple,
  citecolor    = acm-green,
  urlcolor     = acm-darkblue,
  filecolor    = acm-darkblue
}

\newlength{\saveparindent}
\newlength{\saveparskip}
\newcounter{ctr}

\newenvironment{newitemize}{%
\begin{list}{\mbox{}\hspace{5pt}$\bullet$\hfill}{\labelwidth=15pt%
\labelsep=4pt \leftmargin=12pt \topsep=3pt%
\setlength{\listparindent}{\saveparindent}%
\setlength{\parsep}{\saveparskip}%
\setlength{\itemsep}{3pt} }}{\end{list}}

\setcopyright{none}

\begin{document}

\title{Lessons from Trauma-Informed Training on Technology-Facilitated Abuse for Gender-Based Violence Advocates}
\author{Naman Gupta}
\email{n@cs.wisc.edu}
\affiliation{%
    \department{School of Computer, Data \& Information Sciences}
  \institution{University of Wisconsin--Madison}
  \city{Madison}
  \state{Wisconsin}
  \country{USA}
}

\author{Connie W. Chau}
\email{conniewchau@u.northwestern.edu}
\affiliation{
    \institution{Northwestern University}
    \city{Chicago}
    \state{Illinois}
    \country{USA}
}

\author{Sophie Stephenson}
\email{srstephenso2@wisc.edu}
\affiliation{%
\department{School of Computer, Data \& Information Sciences}
  \institution{University of Wisconsin--Madison}
  \city{Madison}
  \state{Wisconsin}
  \country{USA}
}

\author{Kate Walsh}
\email{klwalsh2@wisc.edu}
\affiliation{%
\department{x}
  \institution{University of Wisconsin--Madison}
  \city{Madison}
  \state{Wisconsin}
  \country{USA}
}

\author{Rahul Chatterjee}
\email{rahul.chatterjee@wisc.edu}
\affiliation{%
\department{School of Computer, Data \& Information Sciences}
  \institution{University of Wisconsin--Madison}
  \city{Madison}
  \state{Wisconsin}
  \country{USA}
}

\renewcommand{\shortauthors}{Gupta et al.}

\begin{abstract}


    Technology-facilitated abuse (TFA) is an emerging gendered public health crisis affecting millions of people across the globe. TFA coincides with other forms of gender-based violence (GBV), such as emotional, psychological, and physical abuse by an intimate partner.
    Survivors often seek support from GBV advocates who specialize in helping survivors navigate abusive situations and potential pathways for healing and remediation.
    With the rise in TFA, however, survivors and GBV advocates experience barriers and gaps in knowledge in identifying and mitigating this newer, ever-evolving form of abuse.
    To address this, we developed trauma-informed training as a critical capacity-building intervention to improve GBV advocates' knowledge and skills for responding to survivors and to encourage multi-stakeholder collaboration toward a structural response. We facilitated \ntraining\ training workshops at local, state, and national US-based GBV organizations.
    %
    %
    %
    %
    %
    %
    %
    Our study demonstrates that the trainings increased advocates' perceived understanding of TFA abuse vectors and their impacts on survivors, along with advocates' confidence in providing support, safety planning, and referrals through collaboration with community stakeholders.
    We then conducted retrospective reflections through autoethnography to surface key lessons, contestations, and tensions that emerged from our experiences in designing and coordinating trainings with host GBV organizations.
    %
     Our analysis of autoethnographic notes revealed three key design decisions: coordinating logistics flexibly, reducing reflexive distance from host organizations and attendees, and using our field experience to tailor content for accessibility.
    %
We contribute a concrete checklist to guide future HCI scholars in conducting training interventions for stakeholders in high-stakes contexts.
\end{abstract}

\maketitle

\section{Introduction}
\label{sec:intro}

%

%
Technology has become deeply embedded in intimate partner violence (IPV). Abusers exploit everyday devices and accounts to stalk, monitor, harass, and coerce survivors, practices collectively described as \TFA\ (TFA)~\cite{safetynet_nnedv_technology_2008,safetynet_tech_2021}. Technology is involved in an estimated 63\% of IPV cases, making TFA an increasingly common concern in frontline advocacy~\cite{messing_intersections_2020}.
Survivors commonly seek help from gender-based violence (GBV) advocates~\cite{stephenson_its_2023, gupta_i_2024, stephenson_legal_2025, woodlock_living_2023, flynn_technology-facilitated_2021, flynn_technologyfacilitated_2024}. Advocates understand coercive control and survivor-centered safety planning, but TFA can require them to interpret unfamiliar technologies and ambiguous evidence. Prior work shows that advocates may struggle to determine how abuse occurs or whether a reported threat is technically plausible~\cite{stephenson_its_2023, matthews_stories_2017, freed_digital_2017, flynn_technologyfacilitated_2024}. Limited organizational capacity and the emotional demands of survivor support compound this challenge~\cite{flynn_technologyfacilitated_2024, cuomo_genderbased_2019, tanczer_feel_2021, ramjit_navigating_2024, chau_envisioning_2024}.

Technical information alone is not an adequate response. Disconnecting an account or device may appear to stop an abuser's access, but it can isolate a survivor, reveal survivors' help-seeking practices, or escalate violence~\cite{gupta_i_2024, freed_digital_2017, freed_understanding_2023}. Advocates must therefore interpret technical options through their existing expertise in trauma-informed and survivor-centered care.
%
Advocates may refer survivors to technical specialists or community service providers while remaining attentive to a broader safety plan~\cite{freed_digital_2017, gupta_i_2024, almansoori_web_}. 
%


Trainings offer a way to connect technical and advocacy expertise and is essential for strengthening knowledge and practice~\cite{noauthor_samhsas_2014, purtle_systematic_2020}. Advocates have expressed a desire for formal TFA training~\cite{freed_digital_2017, williams_exploring_2023}. The goal is not to turn advocates into technical specialists, but to help them recognize relevant concerns, assess responses, and seek additional expertise when needed.
%
Unlike a static resource, interactive training allows participants to examine technical guidance through their field experience and consider its safety implications on survivors. It can also build shared language for consultations and referrals. We therefore approach training as collaborative knowledge translation rather than a one-way transfer of information.
This orientation is essential to trauma-informed practice, where trust depends on active listening, empathy, and validation~\cite{wilson_treating_2004, siegel_healing_2003, bellini_sok_2023}. TFA training should reinforce these practices and account for how technological harms intersect with oppression and marginalization~\cite{thompson_embracing_2025, diemer_critical_2016, disalvo_when_2024, ajmani_moving_2025}.

Organizations at global, national, and local levels have begun offering TFA training to advocates~\cite{unfpa_tfgbv_hub, unfpa_tfgbv_programming, ohchr_tfgbv_tool,wesnet_training_2026, nnedv_safetynet_training, nnedv_safetynet_webinars,techsafetycanada_shelter_2026,refuge_tech_abuse_training,esafety_commissioner_au,endtab_cyber_,americanbarassociation_technologyfacilitated_}.
%
%
Yet advocates report that these trainings can lack guidance about appropriate responses in practice or
``really just cover the basics.''~\cite{williams_exploring_2023}.
%
%
More content is not necessarily the solution. Comprehensive coverage becomes outdated as technologies evolve and can burden already over-committed advocates. The design challenge is to select actionable knowledge and adapt it without compromising trauma-informed care~\cite{noauthor_samhsas_2014, chen_trauma-informed_2022, ramjit_traumainformed_2025}.
%
%

Existing programs (e.g. Tech Clinics~\cite{havron_clinical_2019, ramjit_traumainformed_2025}) demonstrate interest in TFA training but offer limited insight into how it is developed with frontline organizations. We know little about adapting a shared curriculum across contexts or evaluating it without burdening advocates. These decisions determine whether training supports or disrupts advocates' work.
TFA response requires distributed expertise and coordination across professional and organizational boundaries. Training is one setting in which advocates and technical specialists can establish a shared understanding of threats and make academic technical knowledge useful in practice~\cite{colusso_translational_2019, yoo_translating_2025, brereton_ethnography_2014, chordia_social_2024}.
Viewing training as cooperative work shifts attention from information delivery to its use in practice. It asks whose knowledge shapes the intervention, how expertise is valued, and how researchers remain accountable to community partners.
Researchers and stakeholders must therefore shape training through a sensitive collaboration. Otherwise, an intervention may be ineffective or cause harm
~\cite{smith_decolonizing_2021, booker_participatory_2016, udah_decolonising_2024, ajmani_whose_2024}.
%
Examining these tensions reveals how technical and care expertise move across professional boundaries, how training adapts to local needs, and how facilitation can seed a reciprocal long-term research partnership. These insights may inform other high-stakes educational interventions.

We investigate these questions through \ntraining\ trauma-informed TFA workshops conducted with \ntotal\ advocates from US-based GBV organizations between July 2025 and June 2026. Drawing on our experience as \techclinic\footnote{More details on Tech Clinics in \autoref{bg:tfa}.} consultants and violence-prevention experts, we developed learning objectives that complemented advocates' expertise with technical knowledge for survivor-centered practice.
We address the following research questions:
\begin{researchquestions}
    \item How do TFA training workshops shape GBV advocates' perceived understanding of TFA and confidence in supporting survivors?
    \item What design, coordination, and knowledge-translation tensions arise when researchers and GBV organizations conduct TFA training in a high-stakes context?
\end{researchquestions}

\smallskip

\noindent We make the following key contributions in HCI/CSCW:


\noindent\textbf{Educational Intervention for TFA:}  We create an educational intervention that embolden US-based GBV advocates with knowledge and agency to identify TFA. Through a mixed-methods analysis, we surface advocates' learning needs related to TFA
and examine changes in their perceived understanding and confidence (\autoref{phase:rq1}). Our findings show how advocates interpreted TFA harms and considered support, safety planning, and referrals. 
The training slides is given in the supplementary material and will made open-source with the publication of this paper.  We believe such an intervention informs crucial scholarship and advocacy towards scalable violence prevention and mitigation in communities.
\smallskip

\noindent\textbf{Collaborative Knowledge Translation:} Through autoethnography, we examine how field experience, positionality, and relationships with host organizations shaped the intervention (\autoref{phase:rq2}). We surface tensions in determining actionable content, adapting training to local contexts, and balancing evaluation with reciprocal service. From these findings, we derive \nlesson\ lessons and \nchecklist\ checklist questions for HCI scholars to create other high-stakes educational interventions in the future.

\section{Background \& Related work}
\label{sec:bg}

We provide a background on the impacts of technology-facilitated abuse on survivors and advocates (\autoref{bg:tfa}) and challenges with trainings in high-stakes contexts (\autoref{bg:hci}).

\subsection{Technology-facilitated Abuse (TFA)}
\label{bg:tfa}

TFA is a growing concern for survivors and GBV advocates. TFA includes spying, stalking, online abuse, hate speech, doxxing, and other forms of harassment using technology~\cite{freed_digital_2017, safetynet_nnedv_technology_2008, chatterjee_spyware_2018, gupta_critical_2024}.
%
%
Abusers use readily available spyware~\cite{roundy_many_2020, chatterjee_spyware_2018, harkin_commodification_2020}, repurpose dual-use applications~\cite{almansoori_global_2022, brookfield_things_2025}, compromise accounts~\cite{afrouz_nature_2021, rogers_technology-facilitated_2022, freed_stalkers_2018, freed_digital_2017},  and abuse smart-home technology like smart speakers or door locks~\cite{stephenson_abuse_2023, slupska_safe_2019, stephenson_its_2023, leitao_anticipating_2019} to coerce and control survivors. Prior research has shown that abusers leverage online ``how-to'' guides for conducting TFA through search engines~\cite{almansoori_web_} and social media forums~\cite{tseng_tools_2020}.
The prevalence of TFA has worsened, especially during the COVID-19 pandemic, showing a dire need for intervention and mitigation~\cite{tseng_digital_2021,safetynet_tech_2021, lyons_experiences_2022}.
%

TFA, especially in the case of IPV, can be complex and high-stakes for the survivors.
TFA manifests in physical and emotional abuse and exposes survivors to traumatic experiences, increasing their risk of developing traumatic stress reactions such as anxiety, hypervigilance, or post-traumatic stress disorder (PTSD) among survivors~\cite{pill_trauma_2017}. Moreover, TFA may disproportionately impact survivors belonging to marginalized identities such as \lgbt, Indigenous, Black and Brown women, or students with a disability etc. who face high rates of violence~\cite{gupta_critical_2024, carlson_we_2026, carlson_attending_2021, gupta_lighting_2025, rosay_violence_2016, blackburn_sexual_2008}.
%

Most survivors are not adequately equipped to cope and mitigate TFA~\cite{freed_digital_2017, matthews_stories_2017}. The complexity and inaccessibility of the user interfaces of modern technology and the ``invisible'' nature of TFA make it difficult to identify and mitigate the abuse~\cite{gupta_analyzing_2023, gallardo_detecting_2022, freed_digital_2017}.
Survivors face severe economic, health, and housing insecurity.
The abuse increases survivor's dependence on their partners for necessities of life: limiting access to medical health services, face utility disconnections, housing instability, and food insecurity~\cite{breiding_economic_2017, vonnessen_economic_2022, vothschrag_advocate_2022}. Survivors may not be technology-savvy or have access to technology expert support providers in their communities~\cite{matthews_stories_2017, gupta_i_2024}. To support survivors, Havron et al.~\cite{havron_clinical_2019,freed_is_2019} introduced a \emph{tech clinic} intervention, which provides trauma-informed technical help. So far there are four such clinics operational in the USA, and a critical resource for survivors in those regions. However, the vast majority of survivors in the USA  still rely on GBV organizations and their advocates to cope with TFA.

GBV advocates may lack expertise and confidence to support a TFA survivor~\cite{freed_stalkers_2018,freed_digital_2017}. As TFA is an emerging topic in social work practice, few advocates are trained on how to handle TFA scenarios. Advocates often self-report low confidence in their ability to manage TFA survivors’ technology concerns~\cite{freed_stalkers_2018, freed_digital_2017, freed_understanding_2023}. Advocates may miss the signs of or mis-conceptualize technology abuse~\cite{dunn_it_2021, freed_digital_2017, gupta_i_2024}.
Trainings should build advocates' confidence in recognizing TFA concerns and applying their knowledge in practice. We surface the design and evaluation of training for advocates that center learning about the nuances of TFA.

\subsection{Training Interventions in High-Stakes Contexts}
\label{bg:hci}

The need to prepare GBV advocates for TFA reflects a broader challenge in professional training: translating specialized knowledge into situated practice without displacing practitioners' existing expertise. We describe a training context as high-stakes when participants must apply unfamiliar knowledge to decisions that can materially affect another person's safety or access to support. In such settings, practitioners often act under uncertainty and within organizational constraints. Training must therefore complement their field experience rather than prescribe responses without regard for context.

HCI researchers have studied how technology can support professional learning in consequential settings. For example, technology-mediated interventions have built capacity among community-based workers~\cite{muke_digital_2020, yadav_sangoshthi_2017, yadav_leap_2019, hui_community_2023, poon_designing_2023, esposito_technology-based_2024}. Simulations have also been used to develop sensitive communication practices among law-enforcement personnel responding to sexual assault~\cite{wang_does_2026, oliva_practical_2010}. This work demonstrates the value of interactive and technology-supported learning. However, it primarily describes the intervention or its outcomes, offering less insight into the collaborative work through which training content is selected, adapted, and facilitated with partner organizations.

This process matters because academic knowledge does not move directly into professional practice. HCI research on knowledge translation shows that research must be interpreted and reshaped before it becomes useful in a new setting~\cite{colusso_translational_2019, yoo_translating_2025, brereton_ethnography_2014, chordia_social_2024}. In a high-stakes training, researchers may contribute technical or scholarly knowledge, while practitioners contribute situated knowledge of clients and organizational practice. Host organizations also shape what can be delivered through their priorities and local constraints. Training is thus a form of cooperative work in which participants negotiate what knowledge is relevant and how it should guide action.

Decisions about training content also reflect relationships of expertise and power. Participatory research calls for practitioners' situated knowledge to shape interventions intended for their communities~\cite{vaughn_participatory_2020, udoewa_introduction_2022, ahmed_practicing_2026}. In practice, however, limited time, organizational capacity, or familiarity with an emerging topic may constrain participation. Researchers must therefore examine whose knowledge informed the curriculum, where participation was limited, and how research activities affected partner organizations. 
Privileging one form of expertise can discount practitioners' knowledge and reproduce epistemic inequities~\cite{ajmani_whose_2024}.

Participant outcomes alone do not explain how these decisions shaped an intervention. Reflexive methods can make visible how facilitators' experiences, positionalities, and relationships influenced the training process~\cite{rapp_autoethnography_2018, kaltenhauser_playing_2024, chang_collaborative_2016}. We therefore use autoethnographic reflections to examine how we combined technical and care expertise, adapted workshops to local contexts, and navigated the boundary between research evaluation and reciprocal service.

\section{Positionality Statement}
\label{methods:positionality}

We are an interdisciplinary team that holds academic positions in a large R1
\univ.
Most authors provide direct support services to IPV and sexual-violence survivors as advocates. Our experiences as researchers, violence practitioners, and educators across gender identities are entwined with our reflections and representations in this paper. Three authors volunteer at \techclinic to provide trauma-informed technical support services and have collectively supported more than 180 survivors. 
One author is a licensed clinical psychologist with decades of experience working clinically with survivors of sexual and relationship violence.
Through this project, we reckon with and give back to the land on which we live and the generations of peoples it has supported. We declined compensation for the trainings, even when it was offered, because we envisioned the trainings from a position of non-extraction, relational accountability, and reciprocity.
All authors deeply respect the knowledge embedded in the \hochunk custodianship of \teejop and recognize their continuing connection to land, water, and community in \teejop.
Finally, some authors are abuse survivors themselves, and their experiences have been instrumental in producing the training material while embodying transformational lenses such as Indigenous restorative justice~\cite{harris_transformative_2006, smith_decolonizing_2021} and trauma-informed ethics of care~\cite{noauthor_samhsas_2014, chen_trauma-informed_2022}.
We discuss how we used these transformative lenses in~\autoref{rq1:goal1}.

Next, we describe the methods and findings for the two phases of our study: \textbf{\textit{Phase I: Conducting Training Workshops (\autoref{phase:rq1})}} and \textbf{\textit{Phase II: Reflections on Training Workshops (\autoref{phase:rq2})}}.

\section{Phase I: Conducting Training Workshops}
\label{phase:rq1}

In total, we conducted \ntraining\ trainings between July 2025 and June 2026,
with one pilot training in December 2024 (no data were collected). The \ntotal\ advocates represented varied cultures, experience levels, and practice backgrounds, and worked in a range of survivor-support roles (e.g., administrative staff, counselors or therapists, sexual-assault nurses, case managers, and law-enforcement officers; see demographics in~\autoref{tab:demographics}).
%

\subsection{Developing the Training Content}
\label{methods:training}
We initially developed the learning objectives from prior TFA research, our experience providing technical support through \techclinic, and our experience in violence prevention. We focused on actionable knowledge and resources that would equip advocates with effective TFA identification and mitigation strategies they could apply in their professional practice to support survivors.  After the pilot, we revised the objectives in response to facilitator reflections, participant questions, and feedback from host organizations. Although advocates did not directly co-design the initial curriculum, their feedback shaped later versions of the content.
The learning objectives (LOs) are as follows:
%
%

\begin{newitemize}
\item[\objective{1}] \textbf{Identify TFA}: Identify common forms of technology-facilitated abuse (TFA).

\item[\objective{2}] \textbf{Technology as a Tool of Abuse}: Explain how abusers exploit technology to perpetrate TFA.

\item[\objective{3}] \textbf{Impacts on Survivors}: Describe how TFA affects survivors' safety, well-being, and daily lives.

\item[\objective{4}] \textbf{Survivor Responses}: Recognize how survivors disclose, cope with, and respond to TFA, and the challenges they encounter.

\item[\objective{5}] \textbf{Safety and Support}: Identify safety strategies, interventions, and resources that can support survivors.
\end{newitemize}

The training content was sectioned into portions for each learning objective.\footnote{Please see the training slides in the supplementary material.}
%
We introduced TFA through Carol's story; Carol is a fictional survivor who faced harassment and stalking by her ex-husband. Using the modified TechSafety Power \& Control Wheel~\cite{safetynet_nnedv_technology_2008}, we explained the most common TFA abuse vectors that survivors face and the similarities between TFA and traditional IPV. We also discussed the difficulty of identifying and mitigating TFA.
We mentioned different types of abusers, common reasons for abuse and their powers: Known person (Intimate partner, friends, neighbors, relatives, police etc.), Informal or formal group online (hired by abuser), Stranger (often anonymous), and Nation State.

Our goal was to utilize transformational lenses to validate their feelings and transform them into strength, resilience, and agency. We made intentional content and engagement to validate the skill-set of advocates. Further, we showed how technologies can also be utilized for good: support and healing. For example, how access to AI technologies can make survivors vulnerable to non-consensual images and deepfakes through websites (fake porn creation tools such as Undress, Candy.ai, iNeedThis, CreatePorn), yet how survivors can utilize AI tools to detect synthetic media and trace their origin and LLM chatbots (like Ruth~\cite{parasolcooperative_ruth_2026}) for trauma-informed support to survivors.
We showed how different vectors may have disproportionately harmful effects on survivors' safety, such as sociopsychological harm and difficulty separating from the abuser. We intentionally discussed how TFA may be perceived differently from more visible forms of abuse, such as physical and sexual violence, and the subtle ways in which coercion is used.

To discuss the impacts of TFA and the barriers clients face in mitigating abuse and seeking support (e.g., invalidation by support providers, dependence on technology for work, insufficient mitigation, and the need to collect legal evidence), we used another story from Chen et al.~\cite{chen_trauma-informed_2022}: Jamie's story. We discussed the challenges advocates may face in providing informed support (e.g., barriers in GBV shelters, technically complex interfaces, limited availability of appropriate resources, and compassion fatigue). We also discussed how survivors may lack access to safe and confidential services and how support providers may disenfranchise survivors through victim-blaming, disbelief, and inappropriate advice.
Importantly, we held space for a discussion on culturally-targeted abuse grounded in lived realities of the Indigenous communities in the context of domestic violence and human/sex trafficking e.g., how gaming and social media platforms have become sites of grooming and recruitment of teenagers.
Finally, we provided a list of online and offline community resources and taught advocates how to refer clients by complementing technical support from community personnel (e.g., a car mechanic, IT support employee, Genius Bar employee, library staff member, or tech-clinic consultant) with emotional validation.

%

\subsection{Recruitment Plan}
\label{methods:recruitment}

We were invited by several GBV organizations at the National-level: National Indigenous Women's Resource Center (NIWRC) and National Network to End Domestic Violence (NNEDV), State-level: \mend, and local shelters such as \arc.
These organizations invited us to participate in a one-off event, a series, or an in-person conference. They provide advocates and survivors with long-term support, hotlines, capacity-building training, resources, and policy advocacy. Some organizations provide culturally sensitive support for marginalized rural and Indigenous peoples.
Host organizations handled the logistics and advertised the training to their employees and partners through mailing lists, websites, and internal announcements.
These organizations house front-line social workers who are experts at supporting survivors experiencing violence.
%
All advocates voluntarily contributed their time and energy to our training workshops, demonstrating their commitment to learning more about TFA and helping their clients.
%
We obtained approval from the \univ Institutional Review Board (IRB), who exempted our research (IRB \#: 2025-0886).

\begin{table*}[!htb]
    \centering\tabfontsize
    \mycaption{Demographics}{Aggregate demographics for the \ndemographics\ (out of \nsurvey) advocates who gave us feedback on our training workshops. The advocates may share multiple categories for Role (s) and Race/Ethnicity. All advocates expressed their Gender identity as Female, with a few expressing \lgbt and 2 spirited pipe carrier.
    }
    \begin{tabular}{lc|lc|lc|lc}
    \toprule
        \multicolumn{2}{c}{\textbf{Work with Survivors}}\vrule 
        &
        \multicolumn{2}{c}{ \textbf{Role (s)}}\vrule 
                    & \multicolumn{2}{c}{\graycell \textbf{Age}}\vrule 
                        & 
                            \multicolumn{2}{c}{\graycell \textbf{Race/Ethnicity}} \\
\midrule        
       \graycell  Yes & \graycell18  &
            Administrative Role$^\dag$ & 9 &
            \graycell 18-24 years & \graycell 1 &         
                     
                         White &  12\\
      \graycell No  & \graycell 5  & Counselor/Therapist$^\oplus$ &  6  & 
                    \graycell 25-34 years & \graycell  6 &         
                              
                             Asian &  2 \\
      \graycell   &
        \graycell  & Case Manager &  3 &
                    \graycell 35-44 years & \graycell 6 &            
                          
                              Black/African American &  1   \\
         \graycell   & \graycell  & Sexual Assault Nurse Examiner
                                          & 3  & \graycell 45-54 years & \graycell 6  
                        &  Indigenous/Native American$^\varnothing$ &  6 \\
          \graycell   & \graycell  &  Community Educator 
                                          & 1   & \graycell 55-64 years   & \graycell 3  &   &  \\   
                                                 \graycell   &
        \graycell   & Not affiliated 
                                          & 1 &
                    \graycell  65+ years & \graycell 1 &   
                        &  \\   
                            
         \bottomrule
    \end{tabular}
    
    \begin{tabular}{lc|lc|lc|lc}
         \toprule
        
        \multicolumn{2}{c}{\textbf{Technology Expertise}}\vrule 
                    & \multicolumn{2}{c}{\graycell \textbf{Educational Degree}}\vrule 
                        & \multicolumn{2}{c}{\textbf{Survivors' Place}}\vrule  &  \multicolumn{2}{c}{\graycell \textbf{Experience}} 
                             \\
\midrule        
            \graycell Moderately knowledgeable & \graycell 19 &    
                Graduate or professional degree & 9 &
                    \graycell Urban & \graycell 12 & >10 years        & 7 \\

            \graycell Very knowledgeable & \graycell 2 &    
               University Bachelors Degree  & 8 & 
                    \graycell Rural & \graycell 7 & 1-3 years & 6\\
        
            \graycell  Slightly knowledgeable & \graycell 2 &    
                Vocational or Similar & 2 & 
                    \graycell Indigenous Reservations & \graycell 5 & <1 year & 3 \\
       
            \graycell & \graycell  &    
                Completed Primary School  &  1 & 
                \graycell Suburban & \graycell 3 & 4-6 years & 4 \\
           \graycell & \graycell &    
                Completed Secondary School  & 1 & 
                 \graycell  & \graycell & 7-10 years & 3 
                 \\
           \graycell & \graycell &    
                Some University but no degree  &  2 & 
                 \graycell  & \graycell &  &  \\   
\bottomrule
\end{tabular}
    \label{tab:demographics}
    \Description{.}
\begin{flushleft}\footnotesize{

    $^\dag$ Roles include Admin asst, Intake clinician, Program Manager, Clinical Supervisor, Director, Director of court services, Supervisor, and, Home visitation manager. \\
    $^\oplus$ Roles include Substance use counselor, Psychotherapist, MAT Recovery Specialist, and, Certified Native American Peer Recovery Services Specialist. \\
    $^\varnothing$ Tribal Nations include Delaware tribe, Red Lake Band of Chippewa Indians, Leech Lake Band of Ojibwe, and, Mississippi Band of Choctaw.
}
\end{flushleft}
\end{table*}

\subsection{Evaluation Strategy}
\label{methods:eval}

\paragraph{Surveys.}
The study follows a pre-/posttest design (common among trauma-informed trainings~\cite{purtle_systematic_2020}) to gauge the effectiveness of our training intervention.
The surveys capture feedback from advocates to understand how likely they are to implement the knowledge from these trainings.
%
%
The pre- and post-training survey questions were developed from a series of core competencies for domestic-violence training programs~\cite{broll_evaluating_2012, etherington_evaluating_2021}. After the training, we asked advocates to complete the post-training survey and sent a reminder email containing the training materials (PowerPoint slides) and our contact information. Thirty days after the training, we sent advocates a follow-up survey to assess the utility of the knowledge shared during the training.
Overall, \nsurvey\ advocates completed our surveys: 44 started the pre-training survey (\nnoconsent\ did not consent; \npre\ consented), \npost\ completed the post-training survey, and \nfollowup\ completed the follow-up survey. Although we had a low completion rate (12\%), it is consistent with prior literature on voluntary trainings~\cite{longCaseStudyAnalysis2009}. To improve survey participation, we offered advocates \$5 (USD) Amazon gift cards for completing the surveys. However, this incentive resulted in only two additional follow-up survey responses (see~\autoref{sec:limitations}).

We customized the core competencies to match the service provisioning landscape around TFA~\cite{tseng_care_2022, havron_clinical_2019, freed_digital_2017}, abuse vectors~\cite{freed_is_2019, stephenson_its_2023, gupta_critical_2024}, and coping and referral strategies~\cite{gupta_i_2024, zou_role_2021, stephenson_legal_2025, gupta_analyzing_2023}.
These competencies cover aspects of GBV  and its impacts (e.g., prevalence, dynamics of abusive relationships, trauma), intervention responses (e.g., disclosure, risk management), reflective practice (e.g., self-care, personal values), professional roles in the workplace (e.g., confidentiality, safety), and collaboration within and between organizations (e.g., making referrals). The surveys measure changes in understanding (e.g. impacts of TFA on community, how abusers use technologies for harm) and confidence (e.g providing support services to survivors experiencing TFA) in supporting a TFA survivor using a likert scale from 1 (Strongly Disagree) to 5 (Strongly Agree) with an option to mark ``prefer not to say'' as 0.
%
We divided the survey items\footnote{Please see the survey questionnaires in the supplementary material.} associated with the five learning objectives (see \autoref{methods:training}) into three evaluation goals:
\goalref{1} {Perceived Understanding of How Abusers Exploit Technology},     
    \goalref{2}  {Perceived Understanding of Impacts on Survivor},
    and \goalref{3}  {Confidence in Providing Support, Safety Planning and Referrals}.
%
%
For each goal, we calculated each participant's mean score, the standard deviation,
and the effect size between post-training and pre-training scores.
Moreover, some questions required qualitative responses from the participants, which were used for the qualitative analysis.
%
%
%
%
The first two authors independently and inductively coded the qualitative responses and in-training quotations to create an initial codebook. They then discussed the codebook collaboratively and resolved areas of agreement and disagreement~\cite{richards_practical_2018}. Finally, we organized the codebook using deductive top-level themes corresponding to each evaluation goal (\autoref{tab:evaluation-goals}).

\paragraph{Safety and Ethical Considerations.}
%
Although GBV advocates are experts, we presented the surveys with sensitivity to participants' emotional, psychological, and digital privacy and safety~\cite{bhalerao_ethics_2022, bellini_sok_2023}.
Additionally, participants may share personal, sensitive, or identifiable information during the study. These identifiers, if leaked, may have downstream safety implications for both advocates, their organizations, and survivors. We anonymized any identifiable information in the survey responses, notes, and stored them in a secure folder accessible only to approved study personnel. The participants' emails were stored separately for payment purposes.
Before the training workshop, we sent out the study details, informed consent form, and the pre-training survey delivered through Qualtrics.

\begin{figure*}[!htbp]
    \centering
     \includegraphics[width=0.9\linewidth]{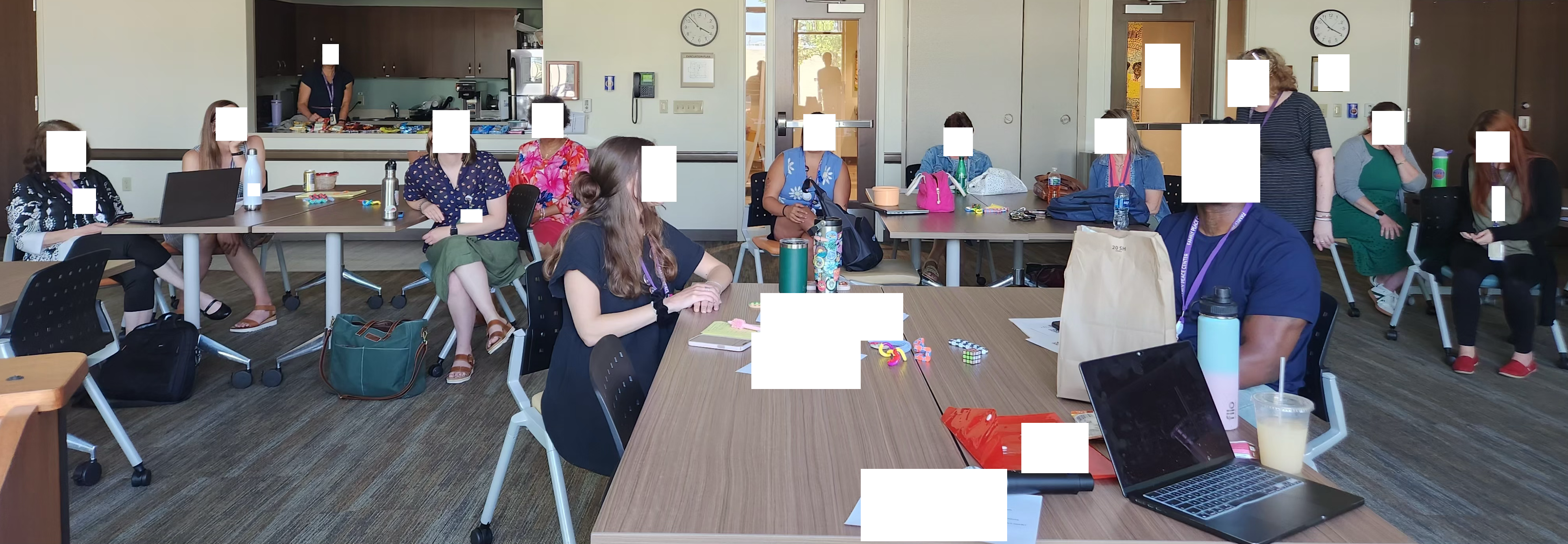}
    \mycaption{In-person Training}{A picture showing advocates participating in-training discussions.}
    \Description{}
    \label{fig:training-pic}
\end{figure*}

\begin{table*}[t]
       \rowcolors{2}{rowAlt}{rowBase}
    \centering
    \tabfontsize
    \mycaption{Usefulness of Learning Objectives}{Advocates rated the usefulness of different parts of the training which covered specific learning outcomes. The advocates were asked to rate the most useful (atmost 3) and least useful (atmost 3), with an option to select more than 1 option or provide a free text response.}
    \begin{tabular}{l|c|c}
        \toprule
        \textbf{Learning Objectives} & \textbf{Most Useful (n)} & \textbf{Least Useful (n)} 
        \\
        \midrule
         \objref{1} {Identify TFA} & 0 & 0\\
         \objref{2} {Abusability of Technology} & 9 & 2 \\
         \objref{3} {Safety Impacts} & 5 & 3 \\
         \objref{4} {Coping Strategies} & 10 & 1 \\
         \objref{5}  {Safety Actions} & 26 & 1 \\
         \midrule
         Prefer not to answer. & 1 & 13 \\
         Something else (responses in prose) & 0 & 10 \\
         \bottomrule 
    \end{tabular}
    \Description{}
    \label{tab:usefulness}
\end{table*}





\subsection{Findings}
\label{rq1:findings}

\begin{table*}[!htb]
\tabfontsize
    \rowcolors{2}{rowAlt}{rowBase}
    \centering
    \tabfontsize
    \mycaption{Effectiveness of Evaluation Goals}{The table provides the pre-training and post-training response means and standard deviation, and the effect size of the trainings.
    }
    \begin{tabular}{
                    m{0.35\textwidth}|
                    m{0.06\textwidth}
                   >{\centering\arraybackslash} m{0.13\textwidth}|
                    m{0.06\textwidth}
                   >{\centering\arraybackslash} m{0.13\textwidth}|
              >{\centering\arraybackslash}      m{0.08\textwidth}}
\toprule
\textbf{Evaluation Goal}  & 
\multicolumn{2}{c|}{\textbf{Pre-Training}} & \multicolumn{2}{c|}{\textbf{Post-Training}}  & \multicolumn{1}{c}{\textbf{Effect Size}}\\
\textbf{(Number of Qs)} & \textbf{N} & 
\textbf{Mean ($\pm$ SD)} & 
\textbf{N} & 
\textbf{Mean ($\pm$ SD)} & 
 \\
\midrule
\goalref{1} Perceived Understanding of TFA Harms  ($n_q=17$) & 51 & $4.212$ ($\pm 0.692$)  & 28 & $4.437$ ($\pm 0.611$)  & 0.338 \\
\goalref{2} Perceived Understanding of Impacts on Safety  ($n_q=9$) & 51 & $4.15$ ($\pm 0.686$)  & 28 & $4.647$ ($\pm 0.586$)  & 0.761 \\
\goalref{3} Confidence in Providing Support, Safety Planning, and Referrals  ($n_q=16$) & 49 & $3.398$ ($\pm 0.659$)  & 25 & $3.982$ ($\pm 0.565$)  & 0.928 \\
\bottomrule
\end{tabular}
    \label{tab:evaluation-goals}
\end{table*}


We discuss preliminary findings on the effectiveness of the trainings.
We use the advocates' quotes from the surveys and in-training discussions. We use an identifier for quotes taken from surveys.
Otherwise, we report the quotes using the generic terms `an advocate', `a few advocates' ($n\le5$), `some advocates' ($6 \le n\le 9$), and `most advocates' ($n\ge10$). We do this since we have a small participant pool and to protect advocates' privacy.
Overall, advocates rated the usefulness of specific learning outcomes. As shown in \autoref{tab:usefulness}, the largest n in the most useful category was for \objective{5} Safety Actions followed by \objective{4} Coping Strategies and \objective{2} Abusability of Technology. A few advocates rated the least useful elements but Safety Impacts and Abusability of Technology were mentioned by a small number.
%
When asked to identify the least useful content, ten advocates instead shared that all parts of the training were useful (e.g., ``All was really informative'',
        ``All was useful. It was great'',
        ``It was all important and [u]seful]'' etc.)

Only \ncommon\ participants completed both pre- and post-surveys.
We conducted paired t-tests with 95\% confidence intervals to determine whether the mean post-test scores differed significantly from the mean pre-test scores [\tstat{\goalref{1}}{-2.860}{0.011},
\tstat{\goalref{2}}{-2.834}{0.012},
\tstat{\goalref{3}}{-4.240}{0.001}].
Moreover, we observed a significant effect size in post-training scores for two of three evaluation goals (see \autoref{tab:evaluation-goals}).
Only for \goalref{3}, the mean of the pre-training scores is statistically significantly smaller than the mean of the post-training scores.
For \goalref{1} and \goalref{2}, although the desired t-test effect may be limited by low statistical power, the pre-training mean responses were closer to Agree (4) than to Strongly Agree (5). This finding suggests that although advocates did not necessarily know the intricate details, they felt confident in their understanding of how TFA manifests and affects survivors.
We also acknowledge that a lack of specificity on what "impacts" or "harm" may mean is open to advocates' interpretation. While we balanced the content on both, we discuss in ~\autoref{sec:limitations} whether the training was better designed for skill-building than knowledge-building. Although interconnected, knowledge can be much harder to shift than behavioral confidence in taking actions. Next, we report themes from individual evaluation goals.




\subsubsection{\goal{1} Perceived Understanding of How Abusers Exploit Technology}
\label{rq1:goal1}
We evaluated advocates' understanding of how abusers use different abuse vectors to harm survivors by helping them ground abusers' capabilities and transform their justifiable skepticism and fear into strength.

\paragraph{Updated perception of abuser capabilities.}
A conversation on abusers' capabilities and abusive power and folk threat modeling stemmed clarifying questions from advocates on nuances of TFA e.g., how abusers can change passwords, how to hide apps, ``An AirPod can work like an AirTag?'', ``Can abusers go to professional hackers?'', ``Capabilities of spyware apps?'', ``How do abusers manipulate smart devices remotely?'', ``Are smartphones are the primary devices that are hacked?'', ``Car used for GPS tracking?''.
Some advocates echoed clients' concerns to check if certain vectors they were facing were even possible or plausible. As one advocate asks \inline{When they call me, they believe the abuser can see where they called and could be hearing the conversation.  Is mirroring a phone real?}
%
%
Some advocates made connections with fictional books and spy movies in Hollywood that depict hacking in a harmful way. We highlighted how much  technical expertise, capability, and physical access to devices that different abuse vectors require, to help separate regular abusers from professional hackers.
Advocates understood how much of TFA stems from abusers' physical access to devices, as P65 shares ``[I learned] that spyware takes a lot of money and physical access to the device in order to implement.''
Further, we helped advocates update their perception of hacking as P4 notes \inline{The risk of "hacking" is far lower than otherwise thought.
}.

\begin{myquote}{P5}
I do feel more comfortable being able to tell them if something is likely, or unlikely to be happening.
\ldots
I learned a lot! It was really helpful learning what someone can actually do, or not do it they do not have access to your phone, and the amount of skill it would require.
\end{myquote}

\paragraph{Transforming fear into agency.}
The content on TFA and the continuously evolving nature of vectors felt overwhelming to the advocates.
When asked about advocates' prior experience when introducing Carol's story (see~\autoref{methods:training}), many advocates echoed the TFA experiences felt as ``scary, magic, powerlessness, creepy, over the top''.
%
Advocates reflected on how TFA can make survivors guarded, an advocate reflects  \inline{TFA brings thoughts in mind of being stalked? Puts you in a different mindset -- like you have to be on guard -- how's he watching me?}
Moreover, advocates raised concerns about emerging technologies (e.g. AI-facilitated porn, IoT surveillance etc), as P67 shared ``[TFA is] scary for the future [with] [n]ew technologies and new systems of abuse''.
We validated advocates' concerns and led intentional conversations to transform their fear into agency.
We shared anonymous survivors experiences from  \techclinic
%
%
and focused on the most-common and easier-to-exploit vectors.

\subsubsection{\goal{2} Perceived Understanding of Impacts on Survivors.}


We evaluated advocates' response to the various impacts of TFA on survivors' overall life, on relationships with children,  family and others, their health, legal options or remedies, and other aspects of survivors' experiences or situations (e.g., mental health, addiction, criminalization).
%
%
Further, we evaluated advocates' understanding of survivors' trauma caused by the abuse, impact on survivors' support systems, and survivors of vulnerable or marginalized identities (e.g., LGBTQ2IA+, immigrant, disabled).
%
%
Advocates affirmed how differently TFA is perceived with other support providers, as an advocate shared \inline{frequently, abusers try to pretend their surveillance comes from a place of concern, and it's really difficult for survivors to convince outsiders it's more than that (eg police, courts).}
%
Further, advocates understood how TFA may come from multiple abusers and cascade from one survivor on to their communities. An advocate shared \inline{Tech abuse can also easily get other people on board with the abuse, through doxing or social media harassment, so the harm can come from multiple people, even if it started coming from just one.}
Advocates reflected on the challenges survivors face, including how invalidation and often-unacknowledged grief can disenfranchise them.
As a forensic nurse examiner shares, ``I've had a few patients over the years who were fearful of someone watching them, some type of video recording in their apartment but police couldn't find anything.  She was easily dismissed as mental health.  Makes me re-think this now. ''
%
Advocates affirmed our culturally-sensitive approach by sharing abuse experiences felt within communities. For e.g. an advocate asked about how \inline{network/phone coverage for survivors/advocates in reservations may impact TFA and location-tracking}.

\subsubsection{\goal{3} Confidence in Providing Support, Safety Planning, and Referrals.}
The last portion of the training covered actions that advocates could take: providing support, developing safety plans, and referring survivors to effective community resources.
%
We evaluated advocates' confidence providing support services to survivors experiencing TFA as an individual advocate, a team, an organization,  and "as a community".
%
%

\paragraph{Validating Survivors' Experiences.}
We evaluated their confidence in creating a safe environment for survivors to disclose TFA, communicate confidentially without unintended impacts, attend to socio-cultural needs (e.g., specific apps, online language, literacy or norms about technology), and if they felt fatigued when providing compassionate care.
We encouraged advocates to leverage their trauma-informed support skills to validate survivors' experiences.
%
%
An advocate shared \inline{I feel more equipped to talk to clients about ways that their technology can be hacked and how to steer them in the right direction for help.} The advocates felt that the conversations could be better supported now that they know how common TFA is, what it looks like, how it can harm survivors, and what the survivor experiences look like.
%
%
%
Finally, advocates responded well to the content focusing on how they can utilize the emotional validation skills that they already have with the intricacies of TFA. P5 reflected \inline{I have always felt comfortable supporting them emotionally and validating etc. and I do feel more comfortable being able to tell them if something is likely, or unlikely to be happening.}



%
%

\paragraph{Mitigating Abuse and Safety Planning.}
Moreover, we evaluated confidence in mitigating the abuse, providing knowledge or resource materials, creating a digital safety plan, aiding the recovery, utilizing the legal remedies available, documenting evidence of TFA in survivor cases. Teaching actionable steps to identify, mitigate, and prevent abuse was a crucial part of the training (e.g. detect hidden apps, spyware, blocking, location tracking, report abuse, two-factor authentication etc.).
Following the practice of trauma-informed trainings, we clarified how some steps to mitigate TFA, which may seem natural to take, may be insufficient, costly, don't work, or escalate violence. A few advocates shared stories of how some actions do not fully mitigate TFA, and therefore, need a nuanced safety plan.

\begin{myquote}{P43}
    We had a Spanish speaking client who limited her tech use to only WhatsApp in trying to get safe from her abuser.  She used it for messages and calls locally and internationally.   It turned out that because she had originally signed up for the app through a phone number on a phone plan with her abuser, he was able to access her content.  It took a long time for this to come to light and has caused many problems for her and many people with whom she communicated.
\end{myquote}

%
%

%

\paragraph{Referrals to Community Resources.}
Finally, we evaluated confidence in
referring survivors to other support providers, and collaborating with the referred providers (people or internal or external organizations).

%
While discussing Jamie's story, an advocate reflected \inline{My first thought is to remind Jamie to connect with someone at the police precinct where here protection order was filed, or a DV/relationship advocate to report her concerns.}. The story prompted discussion about the disenfranchisement and invalidation survivors face within institutions, especially law enforcement.
Nonetheless, advocates felt confident sharing the resources and challenges associated with them, especially, after knowing the abuser capabilities and the possible avenues they exploit. As P3 shared \inline{I can share the information about spyware on phone with them as some of my clients mentioned it and have questions about it. I also love the resources on the slides. I will definitely share those with my clients.}

\begin{myquotel}{P5}
    I feel better knowing I can refer them to someone and to let them know some of the things they suspect, is not possible without access to a phone or without extensive knowledge of technology.
\end{myquotel}

Specifically, the training gave confidence to the advocates (especially, based in \wi) in utilizing the tech clinics, \inline{being able to use you guys [tech clinic] as a resource for our clients is wonderful.} (P5). The training provided a platform to connect advocates and get familiar with the tech clinic services and consultants. As P43 shared, \inline{Meeting you all online gives me more confidence in referring clients to you.  I knew about the clinic from emails but now I can tell them I was in the training and heard more about what can do to help.}
Consequently, a few advocates also made successful referrals and shared information about our tech clinic with clients
, as P3 adds \inline{I shared information of using [Apple] Airtag to track people and provided info about the Tech Clinic for client to reach out.} Despite having access to a tech clinic, advocates were curious in learning the steps they can do during the period before a consultation is scheduled.
Finally, we shared that survivors may be hesitant to seek formal services due to a variety of reasons.
P43 affirmed their clients' hesitance
%
\inline{I have known about your tech clinic and did encourage a client who was going through extreme TFA that we could contact you, however she was very reluctant due to fear of retaliation from the abuser.} We affirmed the advocates' role in validating clients' traumatic experiences and giving them confidence through positive affirmations.

\paragraph{Organizational Readiness \& Workplace Safety.}
We evaluated organizational readiness and workplace safety in the pre-training survey e.g. whether advocates had a clear workplace policy (e.g. specific roles, safety plan, referrals) to address survivors and safety measures to protect advocates from TFA.  Advocates reflected how the training worked as intentional discourse towards their organizational practices.

\begin{myquote}{P1}
This has been a helpful insight when talking about abuse and power/control. It was not a conversation we were having before. This has helped us feel more confident with these conversations.
\end{myquote}

Further, we assessed whether organizations already had training programs about technology or TFA, how to address harms of TFA, understanding support strategies, and safety planning around TFA.
Most advocates (18) reported ``No'' to the question, and some advocates (11) reported that they had a training, with some (14) reporting "Maybe / Not to my knowledge.", and (4) "Prefer not to answer.".
Advocates reported "Neutral" (mean score $~2.88$, $n=32$) for topics covered in the trainings such as
(a) daily use of technology and productivity tools
(b) TFA experiences and their impacts on survivors and other aspects of work
(c) an abuser's tactics to misuse or abuse technology, or
(d) online safety tips and guidance.
%
The below-neutral score underscores the need to embed TFA-related topics in holistic training on technology use and abuse.
When asked how likely are they to recommend this training to other advocates, most   opted `very likely' (24), (7) as `likely', and (1) as `Neither likely or unlikely to recommend'.
This evaluation provides an opportunity to conduct ``an all around advocacy training'' (P37) that integrates knowledge on technology use and abuse. A strong workplace safety policy could help prepare advocates towards violence prevention  (more discussion in~\autoref{discussion:future}).




\section{Phase II: Reflections on Training Workshops}
\label{phase:rq2}


\begin{figure*}[!htbp]
    \centering
    \includegraphics[width=1\linewidth]{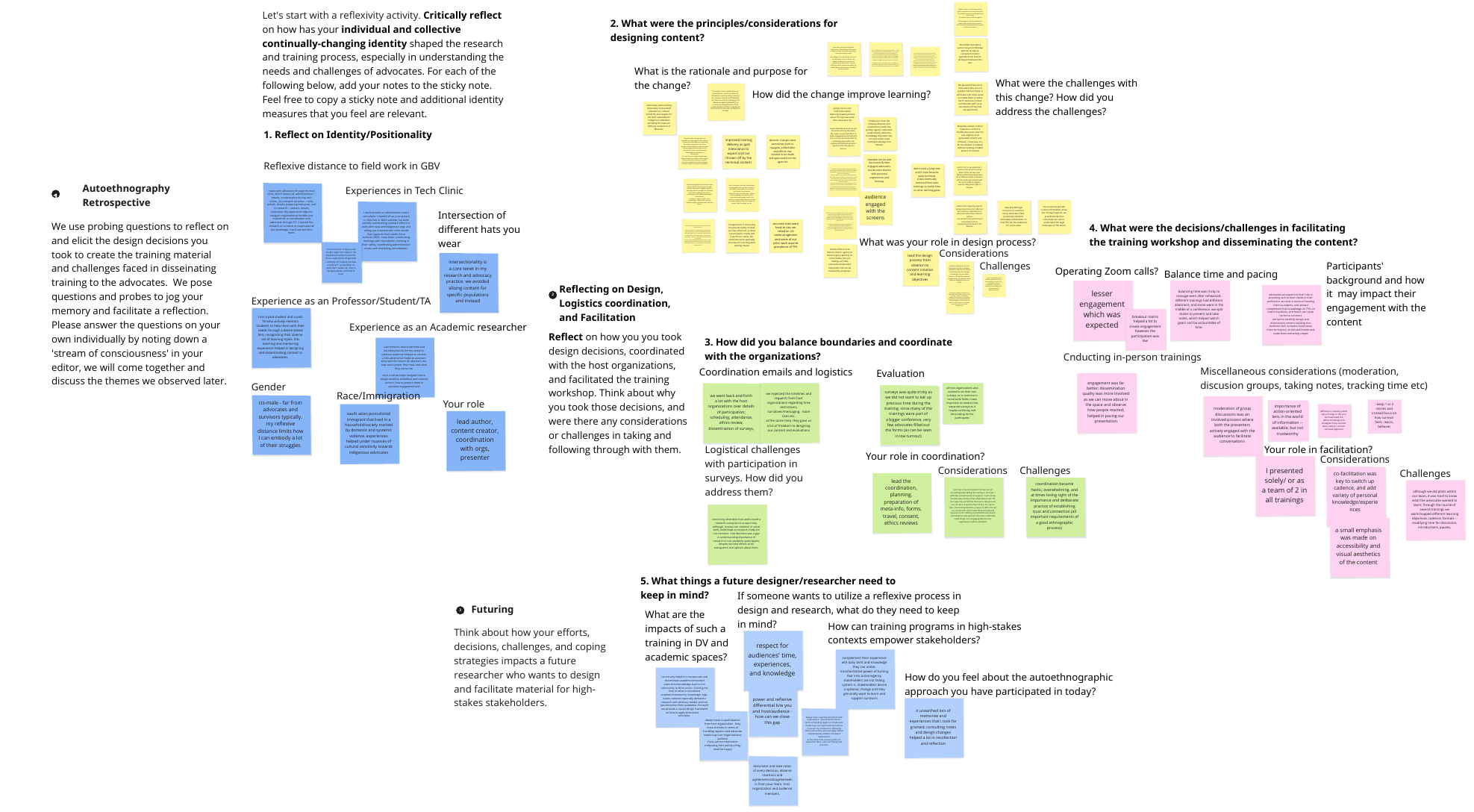}
    \mycaption{Autoethnography Board}{Three authors added reflections through sticky notes on their (1) reflexive identity, (2) principles,  challenges faced and addressed in design, facilitation, and logistical coordination, and (3) how our strategies impacts future researchers in conducting trainings high-stakes contexts.}
    \label{fig:boards}
\end{figure*}

\subsection{Autoethnography}
\label{methods:autoethnography}

We use autoethnography to reflect on, analyze, and synthesize the authors' individual and collaborative experiences and how those experiences affect the research process~\cite{chang_collaborative_2016, sengers_reflective_2006, rapp_autoethnography_2018, kaltenhauser_playing_2024}.
Trainings may be inherently subjective and rely heavily on authors' lived and professional experiences, exposure, and conversations~\cite{cullen_pedagogies_2021}.
%
Because the project spanned a long period (December 2024 to June 2026), we relied on detailed notes from the training workshops and summarized them in a versioned change sheet documenting design changes, challenges, and how we addressed them.
We created a retrospective reflection template Miro board consisting of probe questions (see \autoref{fig:boards}).
The probe questions helped us recall and examine the objective and subjective dimensions of our decisions and collective sense-making.
Three authors added reflections through sticky notes on our (1) reflexive
identity, (2) principles, challenges faced and addressed in design, facilitation, and logistical coordination, and
(3) how our strategies might inform future researchers conducting trainings in high-stakes contexts.
We have added the Autoethnography probe question template in the supplementary material.
After all authors completed their reflections, two authors independently developed a codebook inductively. We discussed the codebook collaboratively and resolved areas of agreement and disagreement. Finally, we identified high-level lessons and takeaways from the authors' retrospective reflections on design, facilitation, and logistical coordination.
%
%

\subsection{Findings}

\begin{table*}[htbp!]
        \rowcolors{2}{rowAlt}{rowBase}
\centering
\tabfontsize
\mycaption{(RQ2) Lessons from Training Workshops}{Takeaways for designing a training \& research intervention.}
\begin{tabular}{m{0.49\textwidth}|m{0.49\textwidth}}
\rowcolor{headerWarm} \color{white}\textbf{Lesson}               
&   \color{white}\textbf{Takeaways from the Lesson}                   \\ 
\toprule
\lessonref{1} Flexibility with Coordinating Logistics with the Host Organizations
&
\bulletpoint Coordinate with Flexibility and Due Diligence. \newline
\\
\lessonref{2} Reduce the Reflexive Distance with Host Organizations and Attendees 
& \bulletpoint Incorporating Advocates Needs and Co-creating Learning Objectives.\newline
\bulletpoint Be sensitive to our shared cultural identity.\newline
\bulletpoint Aligning Motivations to Balance Research and Advocacy.
\\
\lessonref{3} Use Field Experience to Tailor Content for Accessibility.
& 
\bulletpoint Utilize Familiar Language \& Framing.\newline
\bulletpoint Acknowledge Imperfections in Training Content.
\\
\bottomrule
\end{tabular}%
\Description{}
\label{tab:lessons}
\end{table*}

\subsubsection*{\lesson{1} Be Flexible in Coordinating Logistics with Host Organization.}


We followed a structured yet flexible approach to working with host organizations to set up the training workshops. We created a checklist of logistical tasks for each training two weeks in advance and reminded facilitators to complete all preparation requirements. Coordination differed according to the type and location of the host organization, the categories of GBV advocates and services represented, and whether the workshop was a standalone training or part of a larger event (e.g., a conference). All host organizations invited us to conduct the training and co-led logistical planning, advertising, and recruitment.
We communicated with host organizations about participation, scheduling, attendance, ethics review, and survey dissemination. We respected their organizational hierarchies, values, timelines, and requests concerning time restrictions and room sizes. At the same time, hosts gave us substantial freedom to design our content and evaluations. The host organizations were eager to invite an expert-led discussion of TFA, often for the first time in their histories.
%
%
%
Overall, logistical coordination was a time-consuming process that required substantial due diligence. The first author performed most of this work. Having one point of contact may offer benefits, but the role can be taxing. Intentional coordination can also be crucial for building reciprocal relationships with host organizations, which in turn enable long-term evaluation of the trainings and changes in organizational policy and behavior (see~\autoref{sec:discussion}).

\begin{myquotel}{Author}
    Coordination became hectic, overwhelming, and at times losing sight of the importance of slow and deliberate practice of establishing long-term trust and connection, crucial for community-engaged scholarship.
\end{myquotel}


\subsubsection*{\lesson{2} Reduce the Reflexive Distance with Host Organizations and Attendees.}

\paragraph{(a) Incorporating Advocates' Needs and Co-creating Learning Objectives.}
To co-create learning content, we contacted personnel who work in technical roles (e.g., IT support, director of IT etc.) in host organizations for references and potential collaborators.
However, despite our best efforts, our request for collaboration with a full-time advocate, was unfulfilled. Although participation and collaboration were the intentional methodology choices we started with, conversations with stakeholders showed reluctance to collaborate on a topic they have tertiary knowledge. For many advocates it was the first opportunity in their organizations' history to learn about TFA, with some who had \inline{never heard of this} (P14). However, most stakeholders stated their own desire to learn from participating in the trainings and gain experience in order to be able to collaborate on the design in the future.
Consequently, advocates vocalized their needs through survey feedback and in-training questions, which helped shape the broad themes reflected in our learning objectives.
%
Advocates wanted specific steps to identify and mitigate different vectors of abuse. Further, advocates wanted to learn about how to access, utilize and refer clients to the resources in their communities. For eg. P55 asked about resources in light of their clients' prior ineffective practices such as  ``changing your phone or resetting you phone doesn’t always solve the problem.
''.

\begin{myquote}{Author}
    The goals and learning objectives evolved over time and for each training.
    We exercised flexibility in recognizing the needs of groups of attendees.
    We also ensured that we communicated the learning goals beforehand in the description of the emails that were shared with the participants beforehand.
    For example, since our target audience was beyond \wi, we chose to rather spend time on how advocates can utilize the tech-savvy personnel in their communities, instead of a remote consultation with \techclinic that only services survivors in \wi.
\end{myquote}

\paragraph{(b) Being Sensitive to Shared Cultural Identity.}
We felt a moral imperative to make our trainings transformational interventions for combating TFA (see~\goalref{1}). We felt comfortable presenting at certain venues, especially those based in \wi, because we offered direct services to \wi-based advocates.
For larger trainings, we organized breakout rooms and have one facilitator join and mediate conversations in each room. The reduced number of people in the breakout rooms helped with the engagement, although they were not at the same levels of engagement in in-person trainings or when participants knew each other.

\begin{myquotel}{Author}
    Being a White women from \wi felt very comfortable in the broader DV trainings because that was pretty much the audience (White \wi women). But I felt a little more uncomfortable presenting at the \was conference. I felt as if maybe I was intruding by presenting. I also remember feeling that our work was not specific to Indigenous women and was worried about the fit of the presentation for the conference.
\end{myquotel}

Given the reflexive distance between us and the audience, it created a liminal space of discomfort. We utilize the discomforting space to be creative and generative, creating culturally-sensitive content and facilitating with care.
We divided the content into portions that were facilitated by 2-3 authors with diverse identities (except the training at \lco reservation, where only one author was able to travel).
Co-facilitation helped us mix-up voice, cadence and tonality. It was key to embodying our lived experiences and knowledge in the training content.
%
Further, we wanted to return the research and field knowledge we had gained to the community and support advocates in addressing the growing threat of TFA.
%
%
%
%
We ensured that we used culturally appropriate language including land acknowledge with reciprocal intent, abuse vectors rooted in Indigenous communities, and restorative justice approaches to legal interventions.
%
%
Such an intentional framing adapts to immediate needs and long-term capacity building of minority communities and  organizations.


%

\paragraph{(c) Aligning Motivations to Balance Research and Advocacy.}
Our motivations to conduct training changed across the course of the project.
Since the team is made up of senior PhD students, a few authors felt that the training for non-technical field experts would be a good practice for future teaching careers and add to our resumes.
%
Initially, we conducted a pilot training with a \wi-based organization and offered the free \techclinic service, as a means to establish trust and reputation and provide services to rural and Indigenous organizations through a remote tech-clinic service delivery mechanism.
Moreover, as violence experts/ privacy \& researchers, grounding various abuse vectors (e.g. dual-use apps, smart devices, non-consensual image sharing) with abuser power helped in creating a folk mental/threat model.
%
%
We provided actionable knowledge to the advocates that helped us close the reflexive distance and establish trust.

We led the trainings with reciprocity for the community that we benefit from in our programs, while, navigating the boundary between real-world application of filling the need for capacity-building trainings versus academic research.
We were cautious of conducting effectiveness evaluation (e.g. abstaining from pushing surveys, remuneration), not applying methods such as randomized controlled trials (e.g. disseminating knowledge to maximum number of advocates,  engaging in peer-to-peer learning), avoid coercion, and over-burdening the advocates.

\begin{myquote}{Author}
    Convincing attendees how useful could a research survey be to us was tricky. Although, surveys are common in social work, workshops as research study may not be common. I felt like there was a gap in understanding importance of research in non-academic participants, despite our best efforts to be transparent and upfront about them - how the data we collect can be used for driving our careers but also drive policy changes down the line.
\end{myquote}

Many host organizations conducted their own feedback surveys, as a common organizational practice in social work. Thus, conducting two separate surveys for our study and the host organization may have been confusing and time-taking for the participants.
Moreover, we acknowledge that any translation research should provide a safe space for advocates' to voice their needs and contribute to improvements in future trainings and downstream policy interventions around TFA. Therefore, we made best efforts in making sure research was optional and politely requested participants to fill out the post-training surveys as well as organization surveys. We discuss the challenges of research and program evaluations in \autoref{discussion:lessons}.

\subsubsection*{\lesson{3} Use Field Experience to Tailor Content for Accessibility.}

\begin{myquote}{Author}
    We spent lots of time trying to make our first version great. But we couldn't be sure how it would help with advocate learning. We (or at least I) was not working from pedagogical theories, I was merely trying to figure out what needed to go in the talk and how to present it.
\end{myquote}
\paragraph{(a) Using Accessible Language \& Familiar Framing.}
Given that these trainings are the first exposure to TFA for many advocates, we disseminated content with utmost care and responsibility. A few authors attended workshops that taught how-to do anti-violence trainings at conferences. This knowledge was instrumental in capturing the semantics and framing that maybe more familiar to the advocates (e.g. using survivor/client instead of `user').
For example, we modified the TechSafety's Power \& Control Wheel~\cite{safetynet_nnedv_technology_2008} to provide specific examples of technologies exploited by abusers to seed a discussion on folk threat modeling (see more in \goalref{1}).
We used stories to engage in discussion activity, giving a generative space for advocates to share about their personal or experience and reinforce learning and meaning-making through collaboration.
 Stories use a trauma-informed framing to evoke relatable experiences by placing audience members `in her shoes', a common practice for fostering empathy toward victims in violence training~\cite{clevenger_seeing_2017}. Such content grounds technical abuse in the specific harms survivors experience and highlights strategies for coping with abuse. For example, Jamie (adapted from~\cite{chen_trauma-informed_2022}) is a fictional survivor facing TFA scenarios such as unauthorized phone access, misuse of location sharing, and harassment through impersonation. We annotated Jamie's coping strategies, including taking evasive action and seeking support, and discussed the challenges she faced in doing so.
%

\begin{myquote}{Author}
    We added stats to ground the reality of how common some vectors are. Here our goal was to clearly specify how the tech clinic can best help a client facing different kinds of tech abuse, based on our experience where we have not been able to historically support clients facing harassment. However, spying and stalking has been our expertise. Our intent to show that spyware apps are very rare, and instead dual-use apps are more often used, cheaper, and easier to abuse.
\end{myquote}

At one point we had four stories, so we had to remove some stories to balance emotional sensitivity with practical utility and actionable takeaways for the advocates.
Especially, when contentious design changes were made to suit a specific training workshop versus the foundational training as a whole.
E.g. the training at \lco (45 min) discussed culturally appropriate and more common forms of TFA in \lco reservation (e.g. child grooming, sex trafficking, MMIR crisis etc).

\begin{myquote}{Author}
    It felt bad to remove survivor-centered stories because these feel the most real and contain the complexities of real cases. But you can't give a whole training through story alone (at least not in this time frame?). I think we struck a good balance in the end.
\end{myquote}


We used plain, accessible language and included visuals, walkthroughs, and demonstrations of actual technologies and how they work to demystify technological abstractions for advocates.
%
We made conscious efforts to explain abstract technology concepts (e.g., doxing) in accessible language and through tangible examples. For example, we passed around our personal location trackers so that advocates could understand what these devices look like and quickly identify them when next supporting a survivor.

\begin{myquote}{Author}
    TFA vectors, as we realized from our conversations, can be very abstract. Abstractions are very easily processed by computer science and STEM folks, but others can find it challenging from our past conversations and presentations with social workers. The idea was to align all attendees to a common working definition of TFA; conceptualization of TFA is a big barrier reported by our and other prior work in the field.
\end{myquote}



\paragraph{(b) Acknowledging Imperfections.}
We recognize the imperfections in social justice work in high-stakes contexts as perfection may reproduce social and epistemic injustices~\cite{thompson_embracing_2025}. We accepted the limitations in our experiential knowledge and centered our expert experiences in \techclinic. We acknowledged that we will neither be able to collect exhaustive knowledge to keep up with advancements in technology development nor we wanted to flood advocates with ever-growing list of abuse vectors.
\begin{myquote}{Author}
We had to figure out the prioritized content for the specific training. Tweaking which modules were included and which were not. Sometimes this was easy (e.g., we knew remote consults wouldn't be super relevant for advocates outside \wi) but deciding which types of tech abuse were worth covering (especially since there are soooo many) was difficult. I remember we drew on our knowledge about the types of tech abuse that are most relevant to specific populations, and other contextual knowledge. e.g. [Author] drew on his connections and experiences talking to Indigenous advocates here.
\end{myquote}

Instead, our intent was to ensure advocates are well-versed with the class of vectors and information that abusers exploit (e.g. location, address, images etc.) so they can provide instrumental initial care to the survivors.
We recalled survivor stories in \techclinic, we discussed abuse vectors by highlighting their prevalence: from most common vectors (e.g. account compromise, dual-use apps,  non-consensual imagery) to the least (e.g. spyware).
Such a prioritized-content framing helped us turn advocates' perceived fear of technology into agency by focusing on vectors that are the most common for abusers to exploit and technically simpler to identify and mitigate for survivors.

\begin{table*}[!htbp]
        \rowcolors{2}{rowAlt}{rowBase}
\centering
\tabfontsize
\mycaption{Checklist for Researchers}{Guiding Questions for different stages of design, coordination, facilitation, and evaluation of the educational-research intervention.}
\begin{tabular}{>{\centering\arraybackslash}
p{0.2\textwidth}|
m{0.76\textwidth}}
\rowcolor{headerWarm} \color{white}\textbf{Stage} &  \color{white}\textbf{Questions for Future Researchers}                   \\ 
\toprule
\itshape 1. Preparation \& Needs-Assessment
    & 
\begin{numlist}
    \tlevel{1}
    \slevel Is there an evidence-backed need for such training on the topic? 
    \slevel Are there present infrastructures to understand stakeholders' needs? How can you build non-extractive and collaborative infrastructures to understand their needs?
    \slevel How many team members can commit to designing, facilitation, and logistics coordination? How can you clarify their roles and responsibilities to reduce conflict in collaboration?
    \slevel Why do you have the expertise to lead training on the topic? How can you collaboratively build this expertise among the research team? 
    \slevel How can you collaborate with the stakeholders to complement your expertise? What role(s) would the collaborator take?
    \slevel How can you address expectations and reduce burden on the collaborator? 
    \slevel What are the unique challenges of the topics' context? How can you generate sensitivity and address those challenges?
    \slevel How can you identify potential harmful impacts of your training? Are there ways to mitigate it?
    \slevel How can you evaluate the effectiveness of the training without extraction from the communities? 
\end{numlist}
\\ \midrule
 \itshape   2. Initial Engagement with Host Organization
    & 
\begin{numlist}
    \tlevel{2}
    \slevel Is the organization willing to support the training with advertising, recruitment, and logistics?

     \slevel How would you conduct an ethics review with your organization and the host organization? 
    \slevel How does your training support the organizational goals, impacts and existing infrastructure? 
    \slevel How can you seek and incorporate feedback from partner organizations? 
\end{numlist}
\\ \midrule
  \itshape  3. Designing Content
    & 
\begin{numlist}
    \tlevel{3}
     \slevel What conceptual knowledge seems most useful to teach? Are there ways to curate an accurate conceptual knowledge? 
    \slevel Is there a learning theory, training framework, or a pedagogy you can utilize to create the training content? 
    \slevel How can you leave the training content artifact as imperfect and yet, allow room for improvement? 
    \slevel How can you ensure that the content is updated with the contextual and technological realities? How can you sustain updates in the future? 
    \slevel Are there ways you can format training content that is most accessible to the audience? 
    \slevel How can you provide concrete actionable steps that stakeholders can utilize in their professional work?
\end{numlist}
\\
    \midrule
 \itshape   4. Training Facilitation & 
\begin{numlist}
    \tlevel{4}
    \slevel What are the ways you can engage the audience and build collective knowledge? 
    \slevel How can you bring in facilitators with diverse voices and experiences?  How can you incorporate socio-political-cultural sensitivity in training content? 
    \slevel How would you prioritize important concepts and time constraints? 
\end{numlist}
\\ \midrule
 \itshape  5. Post-Training Followup & 
\begin{numlist}
    \tlevel{5}
    \slevel Is there scope of building sustained reciprocity for more future trainings?
    \slevel How can you disseminate knowledge back to the stakeholders, host organizations, and research community? 
\end{numlist}
    \\ \bottomrule
\end{tabular}%
\Description{}
\label{tab:checklist}
\end{table*}

\section{Discussion}
\label{sec:discussion}

In this section, we consider how future HCI scholars can engage with the evaluation findings and retrospective lessons surfaced through this study. This discussion can help researchers create appropriate training interventions for expert stakeholders in their respective high-stakes contexts (\autoref{discussion:lessons}). To this end, we provide a checklist of questions that future scholars can incorporate into their educational interventions (\autoref{tab:checklist}).
Scholars may use this checklist to ground training material, evaluate interventions with contextual sensitivity, and create opportunities to form long-term research partnerships with stakeholders and organizations. Finally, we present our future plans for the training and its adoption (\autoref{discussion:future}) and discuss the study's limitations (\autoref{sec:limitations}).

\subsection{Lessons for Future Trainers}
\label{discussion:lessons}
We surfaced the epistemic gap of knowledge around TFA among GBV advocates through our training workshops.
Given the emergent nature of some topics, stakeholders may not fully understand their learning needs or how those needs fit within their organizations' capacity-building curricula.
Echoing prior work~\cite{ajmani_moving_2025, ajmani_whose_2024, smith_decolonizing_2021, anzaldua_borderlands_2021, cforiest_remediators_2026}, we urge scholars to establish formal needs-assessment and feedback pathways by collaborating with experts who have extensive field experience to bridge the knowledge gap. Although our initial goal was to leverage a range of community-engaged and participatory methods to reduce reflexive distance (e.g., forming a community advisory board, co-designing content, and co-authoring research)~\cite{vaughn_participatory_2020, brydon-miller_action_2012, lu_contamination_2024}, we were unable to find subject-matter experts with whom to collaborate and instead relied on our own experience at \techclinic (see~\lessonref{2}).
Further, designers could distribute responsibilities in advance to avoid conflict within the team and overburdening stakeholders. This distribution can remain flexible while establishing clear roles and responsibilities within the research team and corresponding points of contact at host organizations.
It is imperative to recognize that training content must balance objective facts with subjective examples informed by facilitators' positionalities and values~\cite{rapp_autoethnography_2018, ashcroft_suggested_2025, pillow_confession_2003, mcclearn_reflexivity_2025} and by community needs~\cite{colusso_translational_2019, yoo_translating_2025, costanza-chock_design_2020}. Translational interventions require careful reflection on potential harms and the incorporation of relational accountability, critical humility, cultural sensitivity, and refusal~\cite{tuck_suspending_2009, pillow_dangerous_2010, jonas_designing_2022, tuck_unbecoming_2014, absolon-kaandossiwin-2011, mcdonald_performing_2026}. Design should embrace epistemic subjectivity and bridge mainstream knowledge with local knowledge systems, lived realities, and the experiences of affected populations and designers (\lessonref{2}). Distinguishing research from program evaluation (\lessonref{2}) can also be challenging because each may require different ethics exemptions, approvals, and forms of stakeholder collaboration. A transparent ethics-review process may require data-safety and privacy agreements with host organizations and scholars' institutions~\cite{petterson_conducting_2025}.
%
Moreover, treating stakeholders as experts in their roles and tailoring content help make complex topics substantially easier to communicate (\lessonref{3}). Generative conflict resolution that avoids contentious interference in organizational processes can also benefit long-term reciprocal relationships~\cite{gandu_conflict_2025, ahmed_casecompass_2026, wilner_rooted_2025}.
Additionally, comprehensiveness versus utility for stakeholders is an intentional design choice that designers need to balance.
Because advocates are frontline professionals working in high-stress environments, our training moved away from traditional ``lecture'' one-way style of knowledge production. We informally leaned into pedagogies that prioritize our lived experiences and transformative learning (e.g. trauma-informed care~\cite{thompson_traumainformed_2022}, centering our own and advocates' reflexivity~\cite{ahmadpour_ethics_2025, clevenger_seeing_2017},  discomfort \& care~\cite{cullen_pedagogies_2021}, counter story-telling that grounds indigenous experiences~\cite{solorzano_critical_2002}, sensitivity to Decolonial Indigenous epistemologies~\cite{tuck_decolonization_2012, wong-villacres_decolonizing_2020, freire_pedagogy_2000}, and refusal to participate in modernity/technology~\cite{tuck_unbecoming_2014, tuck_rwords_}.
Future designers may formally rely on learning theories, pedagogies, probes, frameworks to design and evaluate the trainings and their adoption (some examples in \autoref{discussion:future}).
For example, in health contexts involving trauma-exposed individuals, designers could leverage trauma-informed care~\cite{chen_trauma-informed_2022, huo_barriers_2023} or decolonial feminist care~\cite{asherblackdeer_decolonizing_2026, pendse_treatment_2022} to approach discussions sensitively while embodying communal healing and care. A balanced use of strong pedagogical frameworks can improve design, dissemination, and learning.
Scholars can improve training content iteratively and incorporate feedback while accounting for technological advances. For example, rapid advances in digital technology increase exposure to new abuse vectors while also making safety resources obsolete. Scholars should ensure that content responsibly captures the lived experiences of affected populations, mitigation strategies, and information about newer interventions.

\subsection{Future Plans for Our Trainings and Adoption}
\label{discussion:future}

We sought feedback from advocates about what they would like to change or improve in the training content, logistics, or modality. For example, advocates requested additional details about privacy measures beyond those we had already covered.
One of the most dangerous aspects of TFA is that it may extend to advocates' devices and organizational infrastructure.
In the future trainings, we will modify content and evaluation surveys to deliberate over how advocates can inculcate a safety policy for their workplace.
%
Although we use a pre-/post-test design, which is common in trauma-informed training, there is scope to evaluate organizational readiness and the downstream impacts of training within organizations.
\citeauthor{purtle_systematic_2020}~\cite{purtle_systematic_2020}'s review on trauma-informed trainings found that the most meaningful long-term impacts on behavior change and client outcomes happened when the trauma-informed training intervention were followed up by organizational policy changes.
Our future step is to embed the evidence-based training as a part of broader organizational change through implementation science (and adjacent) frameworks that measure long-term adoption~\cite{waddell_leveraging_2024, shin_nasss_2025, mamykina_scaling_2021, lennon_readiness_2017, lyon_bridging_2023, greenhalgh_adoption_2017}. For example, RE-AIM framework~\cite{glasgow_evaluating_1999} provides guidance for public health interventions beyond the research phase: measures, checklist questions, coding guides, and a tool to help plan the long-term adoption, implementation, maintenance, and sustainability. Such organizational change in practice requires collecting administrative data within host organizations e.g., increase in client referrals, increase in detecting TFA through intake form or trauma helplines, etc. However, getting access to such data requires a deep research reciprocal partnership along with transparent and confidential data use and sharing agreements. We plan to form a community advisory board to help co-design not just the training content, but, also understand how we can embed the intervention within the organization. By echoing feminist and decolonial scholarship~\cite{bardzell_towards_2011, costanza-chock_design_2020, kotturi_sustaining_2024, bassin_intervention_2026, dong_evaluating_2026, ahmed_casecompass_2026, sum_youre_2025}, we plan to maintain updated training content (accuracy of new abuse vectors, community resources etc.) and sustained relationships with GBV organizations through the emotional labor of volunteers at \techclinic.

\subsection{Limitations}
\label{sec:limitations}

We discuss the limitations of the study, how they influenced the results, and suggest areas for future research.
First, our findings do not represent all advocates, TFA contexts, or sociocultural and political experiences of advocates and authors. We are researchers based in the US and conducted the trainings in the US, including in a few Indigenous contexts (see \autoref{methods:positionality}).
Regardless, the data quality was sufficient to find themes and patterns where we reached thematic saturation for the qualitative analysis for effectiveness.
Similarly, the findings shared in \autoref{phase:rq2} are based on our subjective autoethnographic reflections. We therefore offer lessons that may help future scholars conduct their own training interventions.

Second, although we tailored the language to match terminology commonly used in GBV contexts, participants may have misinterpreted the evaluation surveys because conceptualizing technology abuse can itself be challenging~\cite{dunn_it_2021, gupta_critical_2024}. Moreover, surveys may not represent a true increase in knowledge, skills, or behavioral change, all of which can be difficult to measure. Although we received positive initial responses, the one-month follow-up period may not be sufficient to evaluate how advocates use the knowledge in practice.
Instead, we only measure improvement in perceived confidence and understanding by treating advocates as trauma experts.
Moreover, the evaluation surveys can introduce ``teaching to the test'' effect in which the ``training heavily emphasizes specific content that is assessed by survey items~\cite{purtle_systematic_2020}.
To address these limitations in the future, we hope to measure behavioral change and use of the knowledge through scenario-based test scores, quizzes, or task/game completion~\cite{brown-mcbride_helping_2020}.
%




\section{Conclusion}
\label{sec:conclusion}

We conducted \ntraining\ trauma-informed workshops to embolden gender-based violence advocates to inform their support provisioning for technology-facilitated abuse survivors. Our preliminary evaluation indicates improvements in advocates' perceived understanding of TFA and confidence in supporting survivors through safety planning and referrals. These findings do not establish changes in professional behavior. Instead, they show the potential value of combining technical knowledge with advocates' existing expertise in trauma-informed and survivor-centered care.
Our reflections further show that training is not a one-way transfer of technical information. It is collaborative work shaped by facilitators' field experience, relationships with host organizations, and participants' situated knowledge. Effective delivery required flexibility in logistical coordination, attention to reflexive distance, and accessible content tailored to local needs. By surfacing these decisions and tensions, we contribute a checklist for researchers developing educational interventions in other high-stakes professional settings. 

\bibliographystyle{ACM-Reference-Format}
\bibliography{bib}

\section*{Appendix}
\label{sec:appendix}

\subsection{Survey Questions}
The survey questions have been uploaded as supplementary material to limit the word count and pages.

\subsection{Training Slides}
The training slides have been uploaded as supplementary material to limit the word count and pages.

\end{document}